\documentclass[12pt]{iopart}

\usepackage{iopams}

\expandafter\let\csname equation*\endcsname\relax

\expandafter\let\csname endequation*\endcsname\relax

\usepackage{amsmath, amssymb}
\usepackage{graphicx}
\usepackage{dcolumn}
\usepackage{bm}

\usepackage{multirow}

\def\GUE{\operatorname{GUE}}
\def\KS{\operatorname{KS}}
\def\WS{\operatorname{WS}}
\def\W{\operatorname{W}}
\def\GOE{\operatorname{GOE}}
\def\Gin{\operatorname{Gin}}
\def\P{\mathbb{P}}
\DeclareMathOperator*{\argmin}{arg\,min}

\usepackage{xcolor}

\begin{document}


\title{{Random matrix statistics} and safety rest areas on interstates in the United States}

\author{J Cai$^1$, J Peca-Medlin$^2$ and Y Wan$^3$}
\address{$^1$ University of Arizona, Tucson, Arizona 85721, USA}
\ead{jiacai@arizona.edu}
\address{$^2$ Department of Mathematics, University of Arizona, Tucson, Arizona 85721, USA}
\ead{johnpeca@math.arizona.edu}
\address{$^3$ Department of Statistics, Rice University, Houston, Texas 77005, USA}
\ead{yunke.wan@rice.edu}




\date{\today}

\begin{abstract}
We analyze physical spacings between locations of safety rest areas on interstates in the United States. We show normalized safety rest area spacings on major interstates exhibit Wigner surmise statistics, which align with the eigenvalue spacings for the Gaussian Unitary Ensemble from random matrix theory as well as the one-dimensional gas interactions via the Coulomb potential. We identify economic and geographic regional traits at the state level that exhibit Poissonian statistics, which become more pronounced with increased geographical obstacles in interstate travel. Other regional filters (e.g., historical or political) produced results that did not diverge substantially from the overall Wigner surmise model. 
\end{abstract}

\maketitle


\section{\label{sec:intro}Introduction and background} 
Random matrices have been used to answer statistical questions since the early history of random matrix theory (RMT){:} Wishart used random matrices to approximate large random sample covariance matrices almost a century ago{, introducing the Wishart ensemble, $\operatorname{W}_m(n)$}  \cite{Wishart}. {(See Section \ref{subsec:note} for definitions of random variables and matrix ensembles used in this paper and other adopted notation.)} Statistical modeling of physical systems using RMT dates back to Wigner's connection of the level spacings of heavy nuclei to the normalized {(to have unit (sample) mean)} spacings between eigenvalues for an unitarily invariant ensemble of random matrices \cite{Wigner}. {The (normalized) sample spacing data (of size $N$), represented by $\tau = \{s_1,\ldots,s_N\}$ where now $\langle \tau \rangle = 1$ (here, $\langle \cdot \rangle$ represents the sample average), can be presented using the \textit{consecutive level spacings} distribution,
\begin{equation}
    P_N(s) = \frac1N \sum_{j=1}^N \delta(s - s_j),
\end{equation}
for $\delta$ the Dirac mass, which is a probability measure that places equal weight at the location of each spacing (with multiplicity); if {$\tau$} is sufficiently random, one can then expect to find a limiting distribution $P(s)$ that satisfies
\begin{equation}\label{eq: limiting dist}
    \lim_{N \to \infty} \int_0^\infty P_N(s) h(s) \ \operatorname{d}s = \int_0^\infty P(s) h(s) \ \operatorname{d}s
\end{equation}
for a sufficiently nice test function $h(s)$ (see \cite{Marklof_2001} and references therein for an expanded discussion). The standard Kolmogorov-Smirnov distance (cf. \eqref{eq:KS}) between the associated probability measures with densities $P$ and $P_N$ for finite $N$ use $h(s) = \mathbf 1_{[0,t)}(s)$ (using the indicator function $\mathbf 1_A(s) = 1$ if $s \in A$ and $\mathbf 1_A(s)=0$ if $s \not\in A$).} 

The particular model that Wigner used for this limiting comparison distribution $P(s)$ is the ($\beta = 2$) \textit{Wigner surmise} (WS) distribution, with density function given by \begin{equation}\label{eq: WS}
    \rho_{\WS}(s) = \frac{32}{\pi^2}s^2 e^{-4 s^2/\pi}, \quad s \ge 0.
\end{equation}
{This model is the exact distribution for the normalized eigenvalue spacings for the $2\times 2$ Gaussian Unitary Ensemble ($\GUE$), a well-studied RMT ensemble that Bohigas, Giannoni, and Schmit observed should occur in the case of ``generic'' chaotic dynamics when the geodesic flow is hyperbolic and time-reversal invariance does not hold \cite{Bohigas_Giannoni_Schmit_1984}. 

It is well known that $\rho_{\WS}$ is a close approximate fit for the {local statistics} asymptotic distribution for the normalized bulk eigenvalue spacings of $\GUE(n)$, {called the \textit{Gaudin distribution} (cf., \cite{Mehta} and references therein for further discussion; see \cite{Deift_Kriecherbauer_McLaughlin_Venakides_Zhou_1999} for a rigorous proof and unfolding procedure to connect $\GUE(n)$ normalized bulk spacings, for }
\begin{equation}
    {\tau \cdot \psi_n = \{\lambda_{j+1}(A) - \lambda_j(A): j = 1,\ldots,n-1; A \sim \GUE(n)\},}
\end{equation}
{where $\lambda_1(A) \le \lambda_2(A) \le \ldots \le \lambda_n(A)$ denote the (real) eigenvalues of the Hermitian matrix $A \sim \GUE(n)$ and $\psi_n$ is a local normalizing factor made with respect to the associated equilibrium measure (or density of states) to ensure $\tau$ has mean $1 + o(1)$). In particular, $\rho_{\WS}$ provides a good approximation for $G'(s)$, where $G(s)$ is the cumulative distribution function (CDF) for the Gaudin distribution, given by}
\begin{equation}
    {G(s) = \sum_{k \ge 2} \frac{(-1)^k}{(k-1)!} \int_{[0,s]^k} \det\left[K_{\sin}(x_i,x_j) \right]_{1 \le i,j \le k} \ dx_2 \cdots dx_k}
\end{equation}
{where $K_{\sin}$ is the sine kernel,}
\begin{equation}
    {K_{\sin}(x,y) = \frac{\sin(\pi(x-y))}{\pi(x-y)}.}
\end{equation}
{While the density $G'(s)$ is given explicitly as the second derivative of the Fredholm determinant of the integral operator with kernel $K_{\sin}$ on $L^2((0,s))$ (see equation (28) in \cite{BBDS06}), the Wigner surmise distribution  provides a sufficient approximation for $G'$ for applications: the Kolomogorov-Smirnov distance between the Gaudin and the Wigner surmise distribution is smaller than $0.005$ \cite{BBDS06,parallel,subway,Mehta,Venker_Schubert_2015}}.} This comparison then enables a particle system to be viewed as one-dimensional gas interactions via the Coulomb potential (i.e., the Dyson gas), which is well-known to align with $\GUE$ statistics \cite{BBDS06,Mehta}. Of particular note for this model, the repulsive forces between neighboring particles  result in the probability two particles are close together is very small. This contrasts with the Poissonian statistics for non-interacting particle systems that can have Poisson clusters, which are known to have normalized spacing statistics equal in distribution to an exponential random variable with unit mean. {This latter fact aligns with the Berry-Tabor conjecture, where $P(s) = e^{-s}$ for $s \ge 0$ in \eqref{eq: limiting dist}, which holds in the case when the corresponding classical dynamics is {regular or} integrable (i.e., the number of constants of motion matches the degrees of freedom)  \cite{Berry_Tabor_1977}}.

Wigner surmise statistics have been exhibited in seemingly surprising places, such as the spacings between the nontrivial zeros on the critical line for the Riemann zeta function (the Montgomery pair-correlation conjecture) \cite{Montgomery} or between the cone photoreceptors inside chicken eyes \cite{chicken}. Transportation systems have served as a popular and ongoing source for other RMT statistical modeling problems, {famously starting with the work of Krb\'alek and \v{S}eba on the spacings between bus arrivals in Cuernavaca \cite{cuernavaca,cuernavaca2}.  For their model, key contributing factors for the ``repulsive'' forces between successive bus arrivals were the privatized bus system and the profit-driven adoption of hiring informants at set checkpoints along the bus route who would inform the driver (and owner of the bus) of the arrival time of the preceding driver; the drivers would then update their driving speed (speed up or slow down) to effect a sufficient gap in an attempt to maximize profits from potential passengers, which thus garners interactions between buses. Other transportation systems that have since been shown to align with Wigner surmise statistics include the spacings between parallel parked cars \cite{parallel}, cars stopped at traffic lights \cite{stopped}, traffic congestion models \cite{superstats}, and arrival times for New York City subway trains \cite{subway}.}

The particular fit to Wigner surmise statistics for these disparate systems aligns with  theoretical and experimental results that spectral properties for coherent chaotic quantum systems appear generic, also known as the universality of quantum chaos (cf. \cite{Haake,KK10}, and references  therein). {This also aligns with the universality phenomenon from RMT. Implications beyond empirical results or observational studies may yet remain elusive for many of these results. For instance, the Montgomery pair conjecture connects intimately to the Riemann hypothesis (RH), which states the non-trivial zeros of the Riemann zeta function lie on the critical line $\operatorname{Re} s = \frac12$. This conjecture, which has substantial supporting empirical evidence as provided by Odlyzko who computed spacings between the first $10^{20}$ non-trivial zeros, provides further credence for the Hilbert-P\`olya conjecture that states the zeros of the Riemann zeta function are the eigenvalues of an unbounded linear operator, and RH is then equivalent to the statement that this operator is self-adjoint \cite{Montgomery,Odlyzko_1987}. If this conjecture were to be proven true, then it would be expected that this (unknown) operator would behave like a random operator \cite{Odlyzko_1987}. In that case, then the distribution of the zeros of the Riemann zeta function should behave asymptotically like those of the eigenvalues of a random Hermitian matrix.}

{In the original Cuernavaca study, the authors additionally studied the empirical number variance, $N(T) = \langle (n(T) - T)^2\rangle$, which measures the fluctuations in the number $n(T)$ of buses that arrived during the time interval $T$, and further measures the spectral rigidity in the model \cite{cuernavaca}. Their results showed a strong alignment with the theoretical GUE asymptotic limiting number variance distribution for only small and moderate $T \le 3$ (cf. Figure 3 from \cite{cuernavaca}), while the normalized spacing statistics showed an overall strong Wigner surmise fit across the entire parameter range. To provide theoretical justification for these observations, Baik, Borodin, Deift and Suidan introduce a microscopic nonintersecting path model for the Cuernavaca bus line that presents itself as a set of independent, rate one, Poisson processes conditioned not to intersect, which leads directly to GUE statistics through standard analysis of orthogonal polynomial ensemble asymptotics \cite{BBDS06}. Of note from their findings and observations, for true GUE statistics one should expect interactions between all particles (such as measured by the number variance) rather than only a three-body interaction or short range interaction (as the normalized bus spacings measure). Subsequent studies present models using cellular automaton \cite{cuernavaca2}, Damped unitary ensemble \cite{Krbalek_Hobza_2016}, and minimizing mutual information from information theory \cite{Warcho_2018} approaches to further support the empirical alignment with the GUE $\beta = 2$ model. However, these approaches do not fully justify \textit{why} the $\beta = 2$ statistics manifest in the empirical data rather than, say, $\beta = 1$ statistics that would follow the Gaussian orthogonal ensemble (GOE) or another $\beta$-ensemble.
}

The main goal of this paper is to extend the transportation systems that exhibit Wigner surmise statistics to include the physical spacings between the locations of safety rest areas on interstates in the United States. Unlike other mentioned transportation systems, which studied the interactions between moving vehicles (e.g., cars, buses, subway trains), we are studying the interactions that result from the budget constraints and public need considerations that went into the planning and construction of stationary   facilities built over multiple decades. {Hence, our aim is to highlight a novel connection between random matrices and safety rest areas through the particular shared {unfolded nearest neighbor spacing} statistics for eigenvalues and site locations. 
}

Construction of the now over 46,000 mile U.S. Interstate Highway System   formally commenced as part of the \textit{Federal-Aid Highway Act of 1956} (Pub. L. 84–627) \footnote{This act is often hailed as a major accomplishment of Eisenhower's two terms (e.g., \cite{Eisenhower,Smith}), although the final form of the bill that was signed into law diverged significantly from the original proposal Eisenhower had submitted one year earlier to Congress that was soundly voted down (e.g.,  Eisenhower promoted the use of bonds rather than the enacted tax model) \cite{Zug}.} (one year after Wigner's results were first published). A major goal for the act was to standardize road design and limit access to road services to better facilitate the transportation of goods, ideas, and people across state lines in a growing post-World War II economy. Part of this standardization included the  construction of safety rest areas to compensate for the more limited access to local facilities. Original funding allocations for these projects (which are ongoing to this day) used a 90/10 split for federal and state budgets, while construction planning was handled locally by states. {Later reductions in federal spending in 1958 then offset the safety rest area construction and maintenance to primarily fall on the states until more funds were eventually allocated through the \textit{Highway Beautification Act of 1965} (Pub. L. 89–285).} 

A variety of factors went into the selection of safety rest area location sites for each state. {As provided in the U.S. Department of Transportation training manual, \textit{Safety Rest Areas: Planning, Locations, and Design} \cite{sra_tm} (see \cite{sra} for a short overview), using input solicited from states that already had successful safety rest area programs as of the late 1970s to then be shared with other states with still nascent or developing programs, such factors for location sites include} expected traffic flow, driver service needs, existing access to facilities, local (geographical, historical or cultural) attractions, and {a recommended}  55 mile average spacings between sites (to match the contemporary national average distance covered in one-hour travel time).  This resulted in site locations spaced so that they were neither too close to other sites or moderately sized cities, which could provide service facilities available through local businesses, and similarly they were not be too far from either. (Interactions between safety rest area locations then also provide insight into the locations of major U.S. cities, which the interstate system was designed to support.) This inherently introduces repulsive and attractive forces  between site locations that we want to compare to the Wigner surmise. {Moreover, the large number of contributing factors for repulsive forces between successive site locations, which would suggest breaking any potential time-reversal invariance properties, should then suggest super-linear repulsions (i.e., GUE statistics) \cite{Bohigas_Giannoni_Schmit_1984}}

{
The paper is organized as follows. The remainder of Section \ref{sec:intro} introduces additional notation and probabilistic background used throughout the document. Section \ref{sec:data} outlines the methodology used for filtering, gathering, and compiling the empirical data of spacings between safety rest areas on U.S. primary interstates, which is normalized (to unit mean) at the individual state levels before being merged into a larger data set for final analysis. Section \ref{sec:KS test} outlines the Kolmogorov-Smirnov test, which is the main analytical tool used for significance testing comparing our empirical data to the theoretical models for the Wigner surmise and Poissonian statistics distributions, along with a convex combination of these two models. Section \ref{sec:results} presents the main findings of our study, where we show normalized spacings between successive safety rest areas located on major U.S. interstates exhibit Wigner surmise statistics. Furthermore, we show Rocky Mountain region spacings exhibit a closer fit to the Poissonian than the Wigner surmise statistics. Sections \ref{sec:future} and \ref{sec:conclusions} outline potential future directions and applications for this study.
}


{\subsection{\label{subsec:note}Notation and probability background}}
{Let $\mathbb F^{n\times m}$ denote the $n\times m$ matrices with entries in $\mathbb F = \mathbb R$ or $\mathbb C$. Let $A_{ij}$ denote the element in row $i$ and column $j$ of $A \in \mathbb F^{n\times m}$. Let $\mathbf I$ denote the identity matrix and $\mathbf 0$ the zero matrix, whose dimensions are explicitly stated if not implicitly obvious.}

{For random variables $X,Y$, let $X \sim Y$ denote that $X$ and $Y$ are equal in distribution. Standard distributions that will be referenced include the standard (real) Gaussian $X \sim N_{\mathbb R}(\mu,\sigma^2)$, with density $f_X(s) = (2\pi \sigma^2)^{-1/2} \exp(-(s-\mu)^2/2\sigma^2)$, the Poisson random variable with parameter $\lambda$, $X \sim \operatorname{Poisson}(\lambda)$, with probability mass function $\P(X = k) = \lambda^k e^{-\lambda}/k!$ for $k = 0,1,2,\ldots$, and the Exponential random variable with parameter $\lambda$, $X \sim \operatorname{Exp}(\lambda)$, with density $f_X(s) = \lambda e^{-\lambda s}$, $s\ge 0$. If $X_i \sim N(0,1)$ are independent and identically distributed (iid), then $(X_1 + i X_2)/\sqrt 2 \sim N_{\mathbb C}(0,1)$ is a standard complex Gaussian random variable.  For $k \in \mathbb N^+$, then the chi-square distribution with $k$ degrees of freedom (df) is given by $\chi^2_k \sim \sum_{j=1}^k X_i^2$ for $X_i \sim N_{\mathbb R}(0,1)$ iid, while the chi distribution with $k$ df is $\chi_k = \sqrt{\chi^2_k}$. Let $\operatorname{Gin}_{\mathbb F}(n,m)$ for $\mathbb F = \mathbb R$ or $\mathbb C$ denote the real or complex Ginibre ensemble of $n \times m$ matrices, where $A \sim \operatorname{Gin}_{\mathbb F}(n,m)$ if $A_{ij} \sim N_{\mathbb F}(0,1)$ are iid.
}

{
Other standard $n \times n$ random ensembles we will utilize include the standard Wishart ensemble, $A \sim \W_m(n)$, where $A \sim BB^T$ is symmetric for $B \sim \Gin_{\mathbb R}(n,m)$ with $n \ge m$; the Gaussian orthogonal ensemble (GOE), $A \sim \GOE(n)$, where $A \sim (B + B^T)/\sqrt 2$ is symmetric for $B \sim \Gin_{\mathbb R}(n,n)$; and the Gaussian unitary ensemble (GUE), $A \sim \GUE(n)$, where $A \sim (B + B^*)/\sqrt 2$ is Hermitian for $B \sim \Gin_{\mathbb C}(n,n)$. More general $\beta$-ensembles are found in RMT, where $\GOE(n)$ and $\GUE(n)$ are the particular instances that align with the $\beta = 1$ and $\beta = 2$ cases. General $2\times 2$ $\beta$-ensembles have normalized (i.e., scaled to unit mean) eigenvalue spacing distributions determined by density
\begin{equation}\label{eq:beta gen}
    f_\beta(s) = C_\beta s^\beta e^{-c_\beta s^2}, \quad s\ge 0, \quad C_\beta = \frac{2\Gamma\left(\frac{\beta+2}2\right)^{\beta+1}}{\Gamma\left(\frac{\beta + 1}2\right)^{\beta + 2}}, c_\beta = \frac{\Gamma\left(\frac{\beta+2}2\right)}{\Gamma\left(\frac{\beta + 1}2\right)^{2}},
\end{equation}
where we note $\rho_{\WS} = f_2$ from \eqref{eq: WS}. (Note  the $\beta = 1$ and $\beta = 2$ models align exactly with the density functions associated to $\chi_{\beta + 1}/\mathbb E \chi_{\beta + 1}$.) While $f_\beta$ exhibits sub-exponential decay as $s \to \infty$ (encapsulating attractive forces between successive eigenvalues), a quick comparison to gauge a potential $\beta$-model match is to estimate the polynomial decay rate as $s\to0$ determined by $s^\beta$; this latter term determines the repulsion between successive eigenvalues, where the $\beta = 1$ and $\beta = 2$ setups correspond to linear and quadratic repulsions. 
}

\begin{figure}[t!]
  \centering
  {%
        \includegraphics[width=0.8\textwidth]{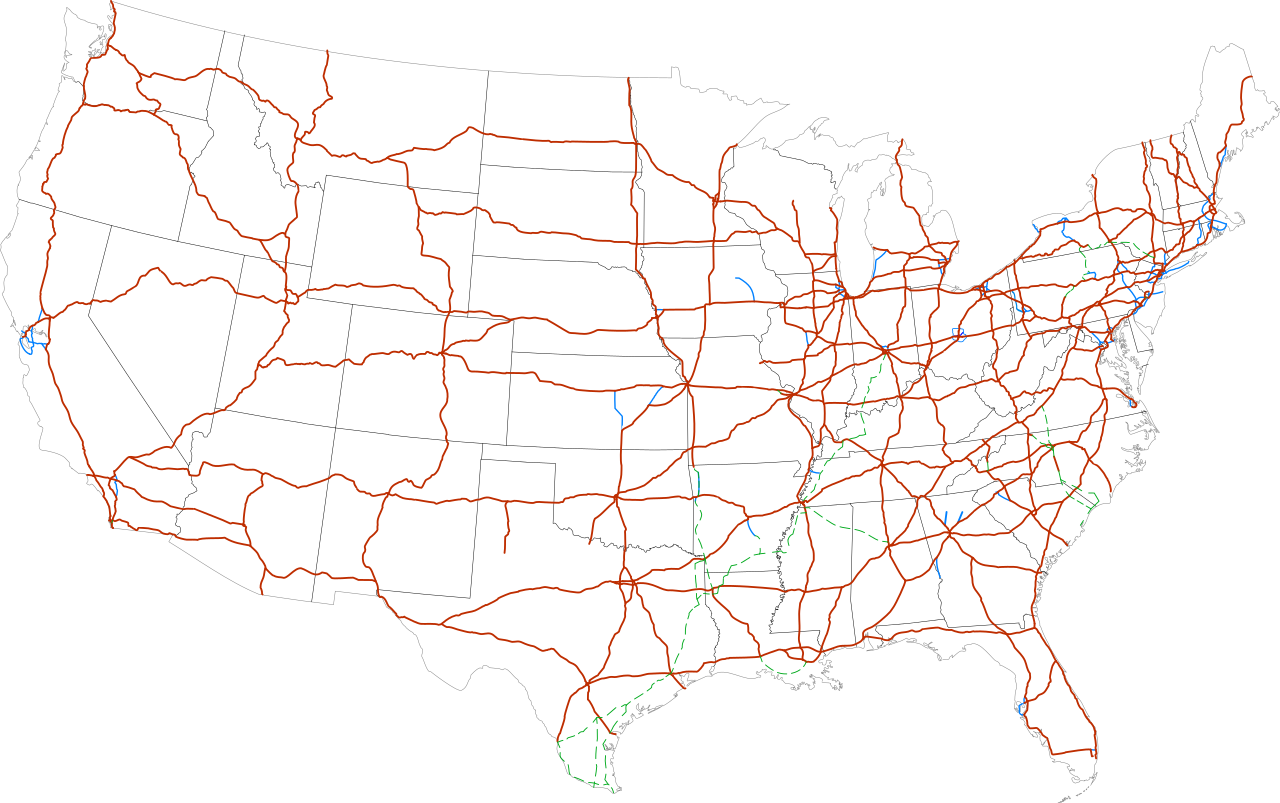}%
        \label{fig:primary_int}
        }%
        \\
    {%
        \includegraphics[width=0.8\textwidth]{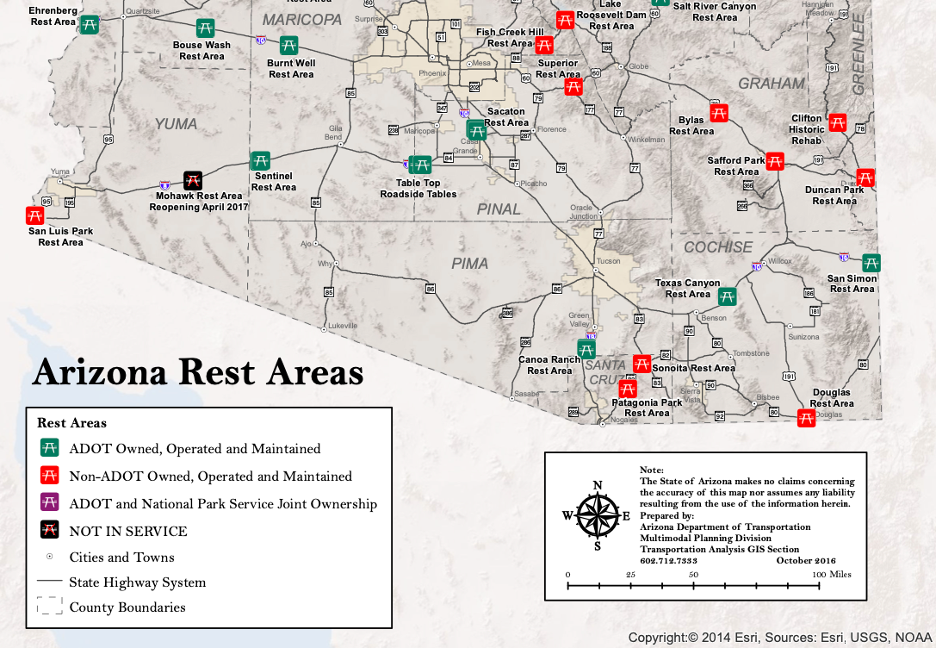}%
        \label{fig:AZ}
        }%
  \caption{Maps of the primary interstates in the contiguous United States (source: Wikimedia) and the safety rest areas in Southern Arizona  (source: AZDOT).}
  \label{fig:interstates}
\end{figure}

\section{\label{sec:data}Data collection} 
Primary interstates comprise the main highways in  the United States  (see FIG. \ref{fig:interstates}) and are labeled of the form I-$N$, where $N$ is a one- to two-digit integer whose parity indicates  the primary cardinal direction alignment of the interstate, where odd numbers (e.g., I-$5$) designate a north-south direction and even numbers (e.g., I-10) designate an east-west direction; this labeling increases in  eastern or northern directions (e.g., I-5 connects the west coast states from California to Washington state and I-95 moves up the eastern corridor from Florida to Maine). Major interstates are numbered by multiples of 5 \footnote{50 and 60 were excluded from interstate naming to avoid confusion with U.S. Highways 50 and 60.}, which typically provide routes that span the United States from coast to coast or from the southern to northern borders (e.g., I-90 stretches over 3,020 miles from Seattle to Boston, while I-8 spans about 350 miles from San Diego to Casa Grande, Arizona). 

For each of the 48 contiguous United States, we use the locations of every eastbound or northbound safety rest area or welcome center located along primary interstates as of June 2023, as provided by each individual state Department of Transportation {(DOT)}. We manually cross-validate each location using the Google Maps API\footnote{Two members of our team independently checked each recorded distance for an additional validation step.} and then record only the segment of the Google Maps computed travel distance in miles on the primary interstate between subsequent site locations (e.g., we do not include any travel distance using  off-ramps or side roads) {when moving in the primary cardinal direction alignment of that interstate (i.e., only eastern or northern directions)}. {Using computed distances provides a higher level of granularity in distances {compared to posted mile post locations found on some state DOT web sites}, which is especially beneficial at later analysis stages due to small or moderate sample sizes. {(We also considered but rejected using the exit number, which typically matches the integral mile distance from the western or southern start point within a state, but in some areas exit number is sequential or even can indicate the kilometer instead of mile distance.)} 
In effect, we project each safety rest area to the closest  point on the interstate to better fit the interpretation of the safety rest areas as particles {located} on the one-dimensional interstate.  We only compute distances between sites on the same interstate, so we do not consider any additional interactions between {different} interstates. FIG. \ref{fig:interstates} shows a snapshot of the safety rest areas located on I-8 (4 stops) and I-10 (6 stops) in southern Arizona in 2016, which matches the available safety rest areas in 2023.


\begin{table}[t!]
\centering
{
\begin{tabular}{|p{1.2in}|m{4.3in}|}
     \hline 
Far West	&	Alaska, California, Hawaii, Nevada, Oregon, Washington	\\ \hline
Great Lakes	&	Illinois, Indiana, Michigan, Ohio, Wisconsin	\\\hline
Mideast	&	Delaware, Maryland, New Jersey, New York, Pennsylvania	\\\hline
New England	&	Connecticut, Maine, Massachusetts, New Hampshire, Rhode Island, Vermont	\\\hline
Plains	&	Iowa, Kansas, Minnesota, Missouri, Nebraska, North Dakota, South Dakota	\\\hline
Rocky Mountain	&	Colorado, Idaho, Montana, Utah, Wyoming\\
\hline
Southeast	&	Alabama, Arkansas, Florida, Georgia, Kentucky, Louisiana, Mississippi, North Carolina, South Carolina, Tennessee, Virginia, West Virginia	\\\hline
Southwest	&	Arizona, New Mexico, Oklahoma, Texas	\\\hline
\end{tabular}
}
\caption{List of Bureau of Economic Analysis regions by state (source: BEA)}
\label{t:econ regions}
\end{table}

For analysis, we filter the data at the state level (e.g., we use a variety of geographical, political, historical, or economical filters, including the Bureau of Economic Analysis (BEA) regions\footnote{BEA regions are statistical areas that are a geographically delineated grouping of states  determined and maintained by the Office of Management and Budget.} (see TABLE \ref{t:econ regions})) or at the interstate level (e.g., only keeping major interstate data). Using the filtered data, {our chosen unfolding procedure uses the}  computed rest area spacings for all interstates within a {\textit{single}} state are merged and then normalized to unit mean by dividing by the sample mean (at the state level). {For example, we define $\hat \tau(s) = \tau(s)/\langle \tau (s)\rangle$ for state $s$, where $\tau(s) = \bigcup_{N \in \mathcal N} \tau(s,N)$ for $\tau(s,N)$ the collection of successive spacings (in miles) between the safety rest areas located on I-$N$ in state $s$ for $\mathcal N$ the filtered set of interstates (e.g., $\tau(\mbox{AZ},10)$ consists of the 5 successive spacing distances (as computed using the Google Maps API for an eastern trajectory) between the 6 safety rest areas located on I-10 in Arizona; cf. FIG. \ref{fig:interstates}).} {No spacings are recorded across state lines.} All normalized state spacings{, $\hat \tau(s)$,} are then merged to form the final test data set, {$\tau = \bigcup_{s \in \mathcal S} \hat \tau(s)$ for $\mathcal S$ the filtered sets of selected states; $\tau$ then also has unit mean {$\langle \tau \rangle = 1$}.}

{Our analysis is thus limited to studying the empirical spacing data between subsequent safety rest areas located on the same U.S. interstate within individual states. Our study does not present other comparison empirical data sets, such as the number variance, since our data set involves stationary locations of facilities rather than moving particles.}

Since individual states are the primary stewards for planning, constructing, and maintaining safety rest areas, using the normalization step at the state level {before merging state data} (rather than at the final step after merging) then has the additional benefit of mitigating any data profile distinctions inherent from different state budgetary constraints.  

We recorded data for 742 safety rest areas on the 70 primary interstates, which resulted in 554 distinct spacing measurements (48.9 mile average (unnormalized) spacing), while the 501 safety rest areas located along the 17 major primary interstates produced 393 spacing measurements (49.4 mile average spacing).


\section{\label{sec:KS test}The Kolmogorov-Smirnov test} The scaled Kolmogorov-Smirnov (KS) test is a non-parametric statistical test that provides a means of comparing two one-dimensional distributions \cite{Kolmogorov,Smirnov}. Hence, this can be used as a fitness test to compare the empirical data set, $\tau$, to a theoretical distribution. We will consider theoretical distributions of random variables with density $\rho(s; u)$ that {continuously} interpolate between independent Wigner surmise and Poissonian statistic distributions, which are of the form  $X_u = (1-u) X + u Y$ for fixed $u \in [0,1]$ and independent $X,Y$ with respective probability densities $\rho_{\WS}$ and $f(s) = e^{-s}${; in particular, $X_0 = X$ and $X_1 = Y$ yield the extreme Wigner surmise (chaotic) and Poissonian (regular) models for our study}. 
{$X_u$ has density \begin{equation}
    \rho(s; u) = \int_0^s \frac{\rho_{\WS}(x/(1-u))}{1-u}\frac{\exp({(x-s)/u})}{u}\ \operatorname{d}x,
\end{equation}
which is the convolution of the densities for $(1-u)X$ and $u Y$
.} This particular form was used in \cite{subway} to profile different {New York City} subway stations, while a similar ansatz was used  in \cite{superstats} to profile different traffic congestion models. {We will carry forward this analysis using our empirical data set of spacings between safety rest areas. This continuous interpolation density $\rho(s,u)$ is analogous to the Berry-Robnik formula for the continuous transition model between the Poissonian and GOE (i.e., $\beta = 1$ Wigner surmise) statistic models used to study semiclassical level spacings when regular and chaotic orbits coexist  \cite{Berry_Robnik_1984}.}

The scaled KS $\alpha$-test statistic we will use is given by 
\begin{equation}
    {c(\alpha) = \sqrt{\#\tau} \operatorname{KS}(\tau,u),}
\end{equation} 
where
\begin{equation}\label{eq:KS}
     \operatorname{KS}(\tau,u) = \sup_{t \in \mathbb R} \left|\frac{\#\{s \in \tau: s \le t\}}{\# \tau} - \int_0^t \rho(s;u) \operatorname{d}s\right|
\end{equation}
denotes the standard KS distance (i.e., the $\ell^\infty$-distance between the {CDF} for each of the empirical and theoretical models) between the consecutive level spacings $\tau$ and $\rho(s;u)$ ($\#\mathcal S$ returns the cardinality of the set $\mathcal S$). {Both compared integrals used in \eqref{eq:KS} match the (non-asymptotic) consecutive level spacing and limiting distribution $P(s) = \rho(s;u)$ found in \eqref{eq: limiting dist} when using the particular test function $h(s) = \mathbf 1_{[0,t)}(s)$.} Standard cut-offs for statistical significance tests for $c(\alpha)$ are the values $1.628, 1.358, 1.224$, which correspond to $\alpha = 0.01,0.05,0.1$, respectively, and align with standard levels of \textit{low}, \textit{moderate}, and \textit{high} statistical significance measurements to reject the null hypothesis that the distributions differ \cite{Kolmogorov,Smirnov}. 

We will also compute the best-fit interpolation parameter 
\begin{equation}
    u^* = \argmin_{u \in [0,1]} \KS(\tau,u)
\end{equation}
for each filtered test data set{, which will serve to assign an effective value of Coulomb potential for different dynamical behavior}. {In particular, $u^*$ will serve as a parameter to help characterize the dynamics as chaotic for small $u^*$, regular for large $u^*$ (near 1), or mixed if sufficiently far from either extreme model; this parameter functions similarly to the use of Brody's repulsion parameter for interpolating around $\beta = 1$ Wigner surmise testing (cf. \cite{Brody,TM85}).} Unlike in \cite{subway}, which had uniform sample sizes for each subway station, we cannot map $u^*$ directly to statistically significant cutoff regions to Wigner surmise or Poissonian statistic fitness since sample sizes can vary widely for each filtered test data set {(cf. Figure 2 from \cite{subway})}. Moreover, for  filters that result in small sample sizes, the statistically significant regions from the $u^*$ mapping to each end point comparison model can overlap.


\begin{figure}[t!]
  \centering
  {%
        \includegraphics[width=0.8\textwidth]{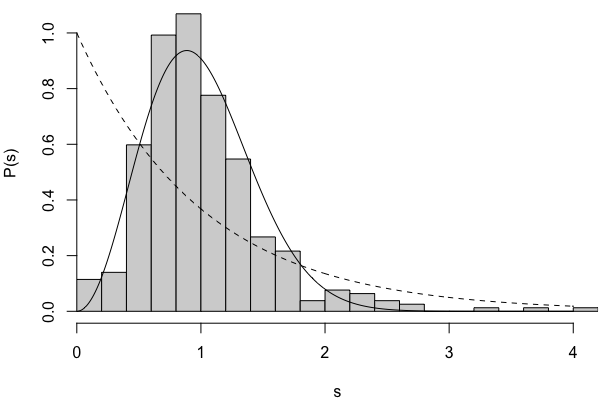}%
        }%
        \\
    {%
        \includegraphics[width=0.8\textwidth]{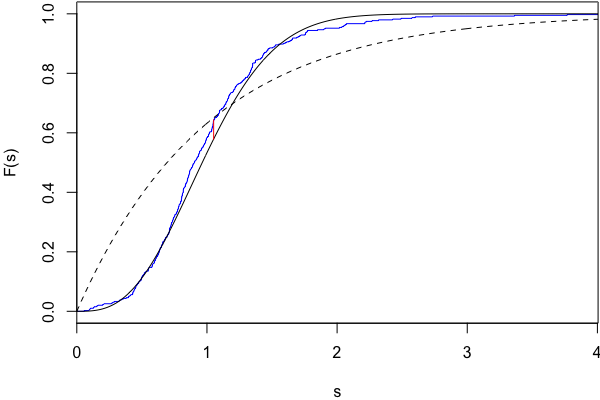}%
        }%
  \caption{Histograms and empirical CDF plot for normalized spacings between safety rest areas on major interstates in the United States. Corresponding (a) density and (b) CDF plots for the Wigner surmise (solid line) and Poissonian statistic (dashed line) models are included for comparison; a vertical line indicates the KS distance between the empirical and WS models.}
  \label{fig:major all}
\end{figure}

\begin{figure}[t!]
  \centering
  {%
        \includegraphics[width=0.8\textwidth]{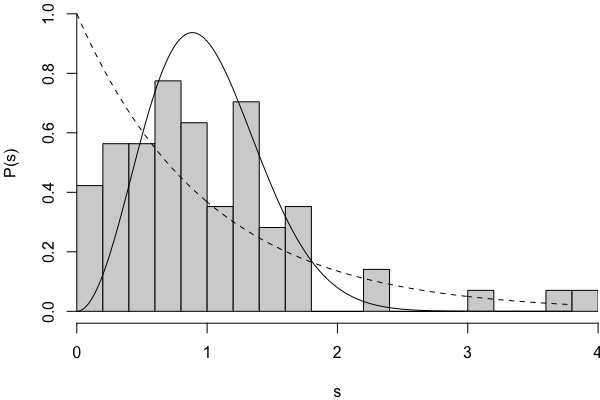}%
        }%
        \\
    {%
        \includegraphics[width=0.8\textwidth]{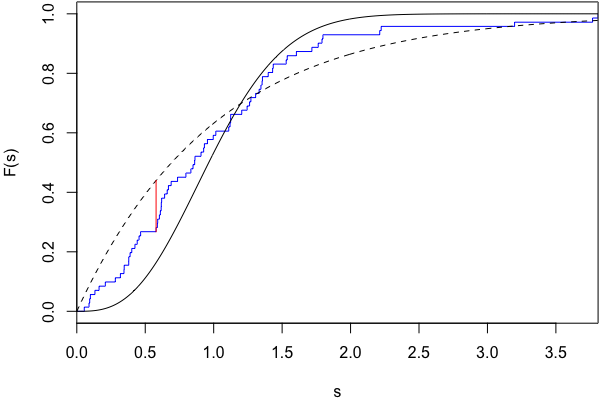}%
        }%
  \caption{Histograms and empirical CDF plot for normalized spacings between safety rest areas on primary interstates in the Rocky Mountain BEA region. Similar comparison lines are maintained as in FIG. \ref{fig:major all}, with now the KS distance in (b) corresponding to the best-fit Poissonian model.}
  \label{fig:rockies}
\end{figure}


\section{Results}\label{sec:results}
We computed the scaled KS $\alpha$-test statistics of our original test data set using  a variety of geographical, political, historical, and economical filters applied at the state level. The computed statistics for the normalized {successive} spacings along all primary interstate and major interstates are included in TABLE \ref{t: summary}. Our main result is we find with moderate statistical significance that normalized successive spacings for safety rest areas located on major interstates pass the $u=0$ KS test (i.e., the Wigner surmise). {FIG. \ref{fig:major all} shows the histogram and empirical CDF for the consecutive level spacings of the normalized safety rest area data mapped against both the ($\beta = 2$) Wigner surmise statistics (i.e., the chaotic model) and the Poissonian statistics (i.e., the regular model). Binning limitations (due to moderate sample size) obstruct visual checks for quadratic repulsions for small $s$ for the histogram versus density function comparisons; the CDF comparisons, however, show a very close alignment for the empirical and Wigner surmise distributions, especially for small $s$.  The CDF plots also show the KS distance between these models (with a vertical red line).}

Most other filters we considered resulted in statistics that did not deviate substantially from the overall model. One notable exception is found in studying the BEA regions. Similar tests show that six of the eight BEA regions pass the $u = 0$ KS test with high significance, while the remaining two BEA regions pass the $u = 0$ KS test with low significance. Of those, only the Rocky Mountain BEA region also passed the $u = 1$ KS test (i.e., Poissonian statistics) with low significance, with a lower $\alpha$-test statistic (i.e., higher confidence) level than the $u = 0$ comparison. {FIG. \ref{fig:rockies}  shows the empirical Rocky Mountain BEA region, again mapped against the Wigner surmise and Poissonian models. Of note, small sample size ($\#\tau = 71$  for Rocky Mountain primary interstate spacings recorded, compared to $\#\tau=393$ used in FIG. \ref{fig:major all} for all major interstate spacings) especially contributes to the small $c(\alpha)$ test statistic. (While other filters manufacture small $c(\alpha)$ test statistics because of small sample size, so some also exhibited statistically significant matching to the Poissonian statistics model, only the Rocky Mountain BEA region of the main filtering methods exhibited \textit{higher} significance matching to the Poissonian model compared against the match to the Wigner surmise models.) Despite the statistical significant match to \textit{both} extreme models, the Rocky Mountain spacings present a strong alignment for intermediate statistics between the chaotic and regular models,  as further indicated by the best-fit interpolation parameter $u^* = 0.737$ (with corresponding test-statistic $c(\alpha) = 0.636$). This was the only regional filter for which $u^* > 0.51${, while the remaining BEA regions had $u^*\in[0.28,0.51]$}.
}

While the BEA regions consider economic similarities between states, it appears the stronger correlation for the Poissonian {or regular} alignment is geographical. The states in the Rocky Mountain BEA region notably have natural geographical obstacles ({viz.}, the Rocky Mountains) that impede travel on interstates, which result in increased travel time (note again our study is comparing \textit{physical} spacings along interstate routes rather than \textit{travel time} {or \textit{arrival}} spacings). For example, a similar analysis shows states with average elevation at least 6000 feet have spacings that fit the Poissonian model ($u=1$) with high significance ($c(\alpha) = 1.053$, $u^*=0.834$), with both Colorado and Utah not only also passing the $u=1$ KS test with high significance but also each having best-fit interpolation parameter $u^* \ge 0.9997$. This mirrors the finding in \cite{subway} that later stations on routes with a higher number of subway stops better align with Poissonian statistics for subway arrivals. In both scenarios, increased obstacles for travel then lead to smaller gaps between locations (for safety rest areas) or arrival times (for subway trains).

Political and historical filters did not yield profiles that diverged as much from the overall picture. For example, we considered the political landscapes at the state level both for today as well as in 1956, immediately after the U.S. Federal-Aid Highway act was passed (see TABLE \ref{t: political}). At the state level for each time period, we record the winners for the most recent presidential election (2020 and 1956 \footnote{Incumbent Dwight D. Eisenhower (Republican) defeated the challenger Alai Stevenson (Democrat) in 1956 while the challenger Joe Biden (Democrat) defeated the incumbent Donald Trump (Republican) in 2020.}) as well as the primary party affiliation for the active governors \footnote{The Democratic-Farmer-Labor party Minnesota governor for both 1956 and 2023 is included with Democrats.}. Despite very different ideological compositions for each the Democratic and Republican parties at each point in time (e.g., 1956 was still in the midst of the transition of the southern Dixiecrat to conservative stalwart \cite{Feldman}), each associated grouping of states maintained a closer Wigner surmise fit. This indicates that the final planning and construction of safety rest areas over many years or decades ultimately remained ambivalent to any potential political influences. Possible implications for this include the deemed public benefit from a healthy and functioning U.S. highway system protected the planning (at the state level) of safety rest areas from any deliberate adverse politicking or the large time scale for planning and construction  mitigated any short-term political finagling between parties.

\vspace{.5pc}

\begin{table}[t]
\begin{tabular}{|r|l|c|ll|c|}
\hline
\multicolumn{2}{|c}{}&&\multicolumn{2}{c|}{$c(\alpha)$} &\\
\multicolumn{2}{|c}{} &&\ \ WS  &\  Poiss.  & \\
\multicolumn{2}{|c}{}&$n$ &($u = 0$) & ($u = 1)$ & $u^*$\\
\hline
\multirow{2}{1in}{Overall}&Primary interstates	&	554	&\		1.733	&\		6.754	&	0.425	\\
&Major interstates	&	393	&\		1.285$^{**}$	&\		5.805	&	0.395	\\ \hline
\multirow{8}{1in}{BEA Regions}&	Southeast	&	128	&\	1.359$^{*}$	& \	3.678	&	0.434	\\
&	Far West	&	45	&\		0.794$^{***}$	&\		2.200	&	0.378	\\
&	Southwest	&	53	&\		0.663$^{***}$	&\		2.200	&	0.507	\\
&	Rocky Mountain	&	71	&\		1.598$^{*}$	&\		1.450$^{*}$	&	0.737	\\
&	New England	&	24	&\		0.731$^{***}$	&\		1.750	&	0.493	\\
&	Mideast	&	54	&\		1.076$^{***}$	&\		2.507	&	0.472	\\
&	Great Lakes	&	90	&\		0.736$^{***}$	&\		3.353	&	0.281	\\
&	Plains	&	89	&\		1.075$^{***}$	&\		3.177	&	0.401	\\\hline
\end{tabular}
\caption{Scaled KS $\alpha$-test statistics for all primary and major interstates and each BEA region with statistical significance indicators ($^*$=low, $^{**}$=moderate, $^{***}$ = high)  and  best-fit interpolation parameter $u^*$.}
\label{t: summary}
\end{table}

\begin{table}[t]
\begin{tabular}{|r|r|l|c|lc|c|}
\hline
\multicolumn{3}{|c}{}&&\multicolumn{2}{c|}{$c(\alpha)$} &\\
\multicolumn{3}{|c}{} &&\ \ WS  &  Poiss.  & \\
\multicolumn{3}{|c}{}&$n$ &($u = 0$) & ($u = 1)$ & $u^*$\\
\hline
\multirow{4}{.4in}{1956} &\multirow{2}{.8in}{Election}&Eisenhower (R)	&	487	&\ 	1.547$^{*}$	& 	6.371	&	0.422	\\
&&Stevenson (D)	&	67	&\ 	1.050$^{***}$	& 	2.443	&	0.493		\\ 
&\multirow{2}{.8in}{Governors}&Democratic	&	339	&\ 	1.685	& 	5.174	&	0.488	\\
&&Republican	&	215	&\ 	0.914$^{***}$	& 	4.591	&	0.372	\\ \hline
\multirow{4}{.4in}{Now} & \multirow{2}{.8in}{Election (2020)}&Biden (D)	&	249	&\	1.097$^{***}$	&	4.478	&	0.439	\\
&&Trump (R)	&	305	&\	1.448$^{*}$	&	5.312	&	0.415	\\
&\multirow{2}{.8in}{Governors}&Democratic	&	255	&\	1.351$^{**}$	&	4.699	&	0.423	\\
&&Republican	&	299	&\	1.456$^{*}$	&	4.907	&	0.458	\\ \hline
\end{tabular}
\caption{Scaled KS $\alpha$-test statistics for 1956 and current political landscapes, including presidential election outcomes and active Governor primary party, with statistical significance indicators and  best-fit interpolation parameter $u^*$.}
\label{t: political}
\end{table}

\section{Future directions of study}\label{sec:future}
A followup study could analyze computed travel times between safety rest areas instead of just the physical spacings along interstate routes. It would be interesting to then reassess the Rocky Mountain BEA region using computed travel times. Some filters we considered result in very small sample sizes, such as for individual states (e.g., Rhode Island had no spacings recorded since no interstate had more than 1 safety rest stop within the state). Another study could focus on higher frequency physical sites located along interstates, such as popular fast food restaurants or gas stations, or just the interstate exits themselves. In the former examples, these methods could be combined with economic performance metrics (e.g., operational key performance indicators (KPIs)) to potentially ascribe desirable or undesirable labels for Wigner surmise versus Poissonian statistic spacing models {(i.e., chaotic versus regular models)}. Our observational study is more limited without available quality metrics for safety rest areas to then classify either spacing statistic model. Similarly, an analogous study could analyze the dynamics underlying other seemingly static networks of facilities (including outside of the United States), such as local access to urgent cares or hospitals, which could additionally then incorporate quality of life or public health KPIs.

\section{Discussion and conclusion}\label{sec:conclusions}
In this paper, we {provide  statistical evidence} to support the statement that the planning and budgetary constraints that went into the placement and construction of safety rest areas along major interstates in the United States resulted in RMT spacing statistics (i.e., the Wigner surmise) that align with the repulsive and attractive forces found in Coulomb gas interacting particle systems. {This suggests a more direct relationship between particular random matrices and safety rest areas, as alluded in the title of this paper. A followup study could further explore interpretations of then the particular random matrix or random operator that then has eigenvalues that match those for the safety rest area locations.} 

While previous transportation systems that  exhibit RMT statistics involved short time scale interactions between moving vehicles, we now expand this set to include a system of stationary site locations planned and built over many decades. {However, our analysis is comparatively more limited since our study focused only on the empirical spacing statistics for our model, while other studies included additional comparison statistics, such as the number variance.} Within this model, we further identify economic and geographic traits at the state level that better align with Poissonian spacing statistics (i.e., the Berry-Tabor conjecture). The Poissonian statistic {or regular dynamics} profile appears to become more pronounced with increased geographical obstacles for interstate travel, while the final placement of site locations appears ultimately independent of any potential political agendas. 

Moreover, using safety rest areas as a proxy for the locations of major U.S. cities along interstate routes, this study yields additional insights into the evolution of the system of cities. In this vein, this study gives merit to using these statistical modeling methods in combination with available quality metrics to analyze dynamics underlying other networks of physical facilities, such as public works or healthcare systems at the national or global scale, which may yield practical benefits and inform public policy discussions.

\section*{References}
\bibliographystyle{iopart-num}
\bibliography{references}

\end{document}